\documentclass[%
 amsmath,amssymb,
 aps,
reprint
]{revtex4-2}

\usepackage{graphicx}% Include figure files
\usepackage{dcolumn}% Align table columns on decimal point
\usepackage{bm}% bold math
\usepackage{hyperref}% add hypertext capabilities
\usepackage{mathtools}
\usepackage{xcolor}

\begin{document}

\title{{Backreaction of perturbations around a stable Light Ring}}

\author{Pedro V.P. Cunha}
 \affiliation{Departamento de Matemática da Universidade de Aveiro and Centre for Research and Development in Mathematics and Applications (CIDMA), Campus de Santiago, 3810-193, Aveiro, Portugal.}
 \email{pvcunha@ua.pt}

\date{\today}% It is always \today, today,

\begin{abstract}

Light rings (LRs) - closed circular orbits of null geodesics - are key features of both black holes and horizonless ultracompact objects. While unstable LRs are relevant for the observation of black hole images, stable LRs have been suspected to trigger instabilities, {namely in exotic compact objects that could mimic black holes}. The underlying mechanism behind this instability remains poorly understood, but a key missing piece is how the {backreaction of a perturbation around the stable LR} modifies the surrounding spacetime. In this work, some progress in this direction is provided by examining a conceptually simple, yet instructive, toy model: continuum-shell stars, supported solely by tangential pressures. Using both analytical and numerical methods, we show how {perturbations around} the stable LR deepen the geodesic potential and shifts its location inward, potentially amplifying any instability associated with the LR. By then extending the analysis to more general stars with nonzero radial pressure, we find that the same phenomenon can be expected to persist under reasonable assumptions.

\end{abstract}

\maketitle
%\tableofcontents
%=========================================================================
%=========================================================================

%%%%%%%%%%%%%%%%%%%%%%%%
\section{Introduction}
%%%%%%%%%%%%%%%%%%%%%%%%

The universe contains extremely dense and massive objects that remain observationally consistent with the black hole paradigm, and cannot be easily explained by conventional stellar models~\cite{Kormendy:1995er,Remillard:2006fc}. However, a fundamental question remains: Are these mysterious objects truly described by the mathematical black hole solutions of General Relativity, or rather by something else entirely different? One intriguing possibility is that black hole candidates are instead described by hypothetical horizonless ultracompact objects which are not black holes, arising from some unknown physics in the extreme regimes of gravity~\cite{Cardoso:2019rvt,Mazur:2001fv}. If such alternatives exist, they could potentially mimic black holes so well that they remain observationally indistinguishable~\cite{Herdeiro:2021lwl,Sengo:2024pwk}. Could such exotic objects be {physically} viable?

When a massive object is sufficiently compact, its strong gravitational field can bend the trajectory of null geodesics into closed circular paths, forming {\it light rings} (LRs). These orbits play a crucial role in both theoretical and observational aspects of black hole physics. In particular, LRs are essential in determining the black hole {\it shadow} image, which is one of the main scientific targets of the Event Horizon Telescope observations~\cite{EventHorizonTelescope:2019dse,EventHorizonTelescope:2019ths,EventHorizonTelescope:2022wkp,EventHorizonTelescope:2022xqj}. In the gravitational wave (GW) channel, LRs have also been associated with key signatures in the ringdown phase of a perturbed black hole~\cite{goebel1972comments,Cardoso:2016rao,McWilliams:2018ztb,Volkel:2022khh}, although some studies have suggested limitations to this connection~\cite{Khanna:2016yow,Konoplya:2017wot}. However, LRs are not unique to black holes. In fact, they are a ubiquitous feature of highly compact objects and can also exist around horizonless objects, $i.e.$ those without an event horizon ~\cite{Glampedakis:2017cgd}.

LRs can be classified based on their orbital stability under small perturbations. In black hole spacetimes, LRs are typically unstable~\cite{Cunha:2020azh}, and these unstable orbits play a crucial role in shaping the observable properties of black holes. For this reason, they are often referred to as ``standard'' LRs. However, in a four-dimensional asymptotically flat spacetime describing an equilibrium object without a horizon, non-degenerate LRs must appear in pairs: one being {\it unstable} and the other {\it stable}~\cite{Cunha:2017qtt,Cunha:2020azh}. The unstable LR behaves similarly to those found around black holes and could allow the object to mimic some black hole-like features. In contrast, the second LR is stable and more exotic. The existence of a stable LR is particularly intriguing, as it has been linked to potential spacetime instabilities~\cite{Keir:2014oka}. Such a LR instability was explicitly observed in fully non-linear numerical simulations of well-motivated horizonless models~\cite{Cunha:2022gde}. Specifically, the presence of a stable LR was found to trigger the instability, causing the object to either i) evolve into a new configuration without LRs or ii) collapse into a black hole. Despite the progress in~\cite{Cunha:2022gde}, the instability was observed in a specific class of models. If such instabilities are generic, they could pose a fundamental challenge to the viability of exotic compact objects as black hole mimickers. However, the underlying mechanism driving the LR instability is still poorly understood. 

A recent breakthrough on this issue was presented in~\cite{Benomio:2024lev}, wherein a non-linear scalar wave was investigated as a simplified model for gravitational perturbations. By numerically solving the non-linear wave equation on top of a static, spherically symmetric, and asymptotically flat spacetime with a stable LR, the study in~\cite{Benomio:2024lev} reported a growth in local higher-order metric derivatives near the stable LR. This growth was driven by a direct energy cascade, transferring energy to increasingly higher-order modes {(see also~\cite{Redondo-Yuste:2025hlv})}. While such a cascade could signal the onset of a broader instability, the analysis in~\cite{Benomio:2024lev} did not explore its backreaction on the stable LR itself. Intuitively, the accumulation of additional energy near the stable LR could backreact on the spacetime geometry, further amplifying gravitational perturbations. Since these perturbations decay slowly around the stable LR, they could initiate a positive feedback loop, fueling an instability. Thus, to uncover the mechanism behind the LR instability, it is crucial to examine {how an energy buildup around the stable LR affects its properties.} {A significant analysis related to this issue was reported in}~\cite{DiFilippo:2024poc}, which explored self-gravitating {\it unstable} LRs around a Schwarzschild black hole. In that work, the authors considered a scenario in which an advanced civilization could artificially populate the black hole's unstable LR with a large number of photons. As the photon population grows, its collective mass becomes significant enough to modify the surrounding spacetime through its {backreaction on the geometry}. This accumulation of photon energy leads to the formation of additional LRs. However, these newly formed structures are dynamically unstable, with even minor perturbations causing the entire configuration to collapse.

In this work, we shall examine how the {backreaction of perturbations around a} {\it stable} LR influences the structure of a {\it horizonless} spacetime. The results suggest that { the existence of additional energy around} a stable LR generally {\it deepens} the geodesic potential well and shifts the LR {\it inward}, potentially amplifying the instability associated with stable LRs. To explore this effect in a controlled setting, we primarily focus on a simple, yet insightful, family of toy models known as {\it continuum-shell stars}. These models provide an analytically tractable framework to isolate and examine the impact of {perturbations around the stable LR}. We introduce this spacetime family in Secs.~\ref{sec:multiple-shells} and~\ref{sec:continuum-shell-stars}. Next, in Sec.~\ref{sec:small_perturbationsLR}, we derive {small perturbative effects coming from the additional energy around} the stable LR, within the continuum-shell star framework. In Sec.~\ref{sec:Non-linear}, we apply the analysis to concrete models that illustrate the consequences of {\it non-linear} perturbations induced by the stable LR. Finally, in Sec.~\ref{sec:beyond}, we attempt to generalize these results to more generic spacetimes, leading to the final discussion and conclusions in Sec.~\ref{sec:conclusions}.  Natural units $G=1=c$ {and a spacetime signature $(-+++)$ are assumed throughout the paper}.

%%%%%%%%%%%%%%%%%%%%%%%%
\section{Static spacetime with multiple shells}
\label{sec:multiple-shells}
%%%%%%%%%%%%%%%%%%%%%%%%

We begin by considering a static, spherically symmetric horizonless star composed of $n+1$ infinitesimally thin matter shells. Each shell is indexed by $k\in \{0,..., n\}$ and positioned at a radius $r_k = r_n\,k/n$, where $r_n$ is the largest shells radius.

Assuming that the regions between these shells are described by vacuum solutions of General Relativity, Birkhoff's theorem~\cite{birkhoff1927relativity} dictates that each region is described by the Schwarzschild metric. Consequently, the metric in the $k$-th region, lying between shells indexed by $k$ and $k+1$, takes the form~\cite{Berry:2022zwv, DiFilippo:2024poc}:  
%================
\begin{equation}
    ds_k^2= -\zeta_k\,\left(1-\frac{2m_k}{r}\right)\,dt^2 + \frac{dr^2}{1-2m_k/r}\, + r^2\,d\Omega^2\,.
\end{equation}

The $k$-th region mass $m_k$ is (essentially) arbitrary, as long as it satisfies $r_k>2m_k$. By requiring continuity of the metric across adjacent regions, the coefficients $\zeta_k$ then satisfy the recurrence relation:
%================
\begin{equation}
    \zeta_k = \left(\frac{r_k-2m_{k-1}}{r_k-2m_k}\right)\, \zeta_{k-1}\,,
    \label{eq:zeta1}
\end{equation}
where $k \in \{1,...,n\}$.
Imposing asymptotic flatness, we normalize the time-like Killing vector $\partial_t$ at infinity, implying $\zeta_n=1$. This condition uniquely determines all remaining values of $\zeta_k$, particularly $\zeta_0$. By introducing the mass of the $k$-th shell as $\delta m_k \equiv m_k-m_{k-1}$, equation~\eqref{eq:zeta1} then leads to:
%================
\begin{equation}
    \zeta_j = \zeta_0\, \prod_{k=1}^j \left(1+\frac{2\,\delta m_{k}}{r_k-2m_k}\right)\,,
    \label{eq:zeta2}
\end{equation}
with $j\in\{0,...,n\}$. An important observation, which will be relevant in the following discussion, is that each matter shell is {\it self-sufficient} in resisting gravitational collapse through tangential stresses: since the regions between shells contain no matter, there cannot be any support in the form of radial stresses from those regions.

%%%%%%%%%%%%%%%%%%%%%%%%
\section{continuum-shell stars}
\label{sec:continuum-shell-stars}
%%%%%%%%%%%%%%%%%%%%%%%%

\begin{figure}[ht]
\includegraphics[width=0.25\textwidth]{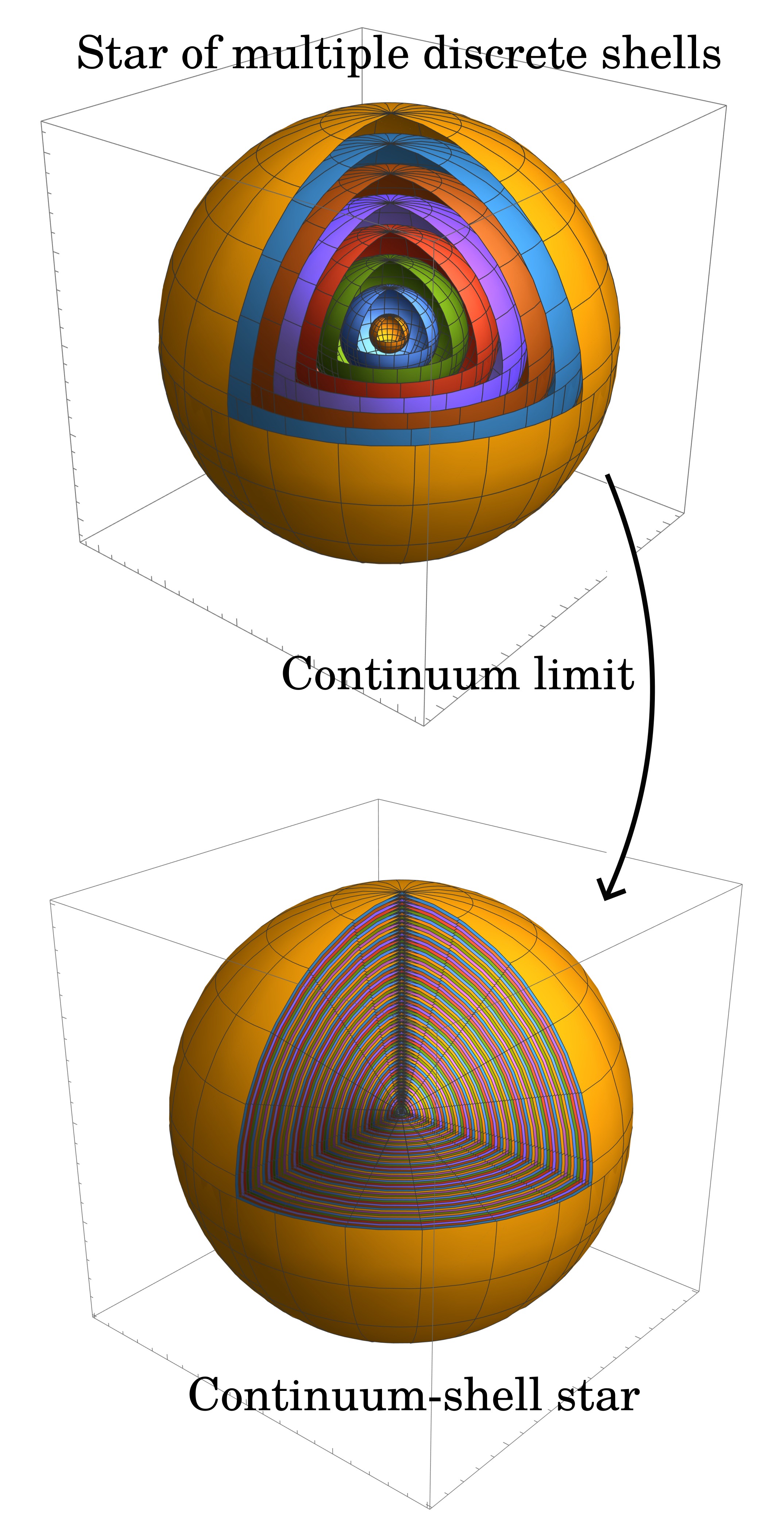}
\caption{\label{fig:illustration} Illustrative image of how a structure made of numerous spherical shells transforms into a {\it continuum-shell star}, in the limit of an infinite number of shells.}
\end{figure}

A particularly interesting case arises when the number of matter shells increases indefinitely while keeping the total star mass $M$ finite. In this {\it continuum limit}, as both $n\to \infty$ and $r_n\to \infty$, the individual masses $m_k$ are promoted to a mass function $m(r)$. As discussed in~\cite{DiFilippo:2024poc}, the continuous limit of the product~\eqref{eq:zeta2} is provided by a Volterra-type integral~\cite{Slavik2007}.
Consequently, equation~\eqref{eq:zeta2} transforms into:  
%================
\begin{equation}
    \zeta(r) = \zeta(0)\exp\left\{ \int_0^r \frac{2m'}{\bar{r}-2m}\,d\bar{r}\right\}\,,
    \label{eq:zetaCont}
\end{equation}
where the prime denotes differentiation with respect to function argument. Since $\zeta_{\infty}=1$, this implies:
%================
\begin{equation}
    \zeta(r) = \exp\left\{ -\int_r^\infty \frac{2m'}{\bar{r}-2m}\,d\bar{r}\right\}=e^{-\psi(r)}\,,
    \label{eq:zetaFinal}
\end{equation}
where the $\psi$ potential is given explicitly by:
%================
\begin{equation}
    \psi(r) \equiv \int_r^\infty \frac{2\,m'(\bar{r})}{\bar{r}-2\,m(\bar{r})}\,d\bar{r}\,.
    \label{eq:psiEq}
\end{equation}
In the continuum-shell limit the metric of the spacetime becomes:
%================
\begin{equation}
    ds^2 = -e^{-\psi(r)}\left(1-\frac{2m(r)}{r}\right)\,dt^2 + \frac{dr^2}{1-2m(r)/r}+r^2d\Omega^2\,.
    \label{eq:metric}
\end{equation}
This spacetime describes a {\it continuum-shell star}: a structure composed of infinitely many adjacent matter shells, see Fig.~\ref{fig:illustration}. See also~\cite{Jampolski:2023xwh}, where a similar (albeit different) concept was recently discussed.

%###########################
%###########################

One of the features of this model is that the spacetime depends solely on a single function: $m(r)$, which can be freely specified as long as $r>2m(r)$. The energy-momentum tensor, derived via Einstein's field equations from the metric~\eqref{eq:metric} in Schwarzschild-like coordinates \( (t,r,\theta,\phi) \), is:  
%================
\begin{equation}
    T^\mu_\nu = \textrm{diag}(-\rho,\,p,\,p_T,\,p_T)\,,
\end{equation}
where $\rho$ represents the matter density, $p$ the radial pressure, and $p_T$ the tangential pressure. In terms of the metric, these quantities are given by:  
%================
\begin{equation}
    \rho = \frac{m'}{4\pi r^2}\,,\, \qquad p=0\,,\qquad p_T= \frac{m\,m'}{8\pi r^2 (r-2m)}\,.
\end{equation}
Notably, the radial pressure $p$ is {\it identically zero}, which is expected, as the spacetime arises from the limit of discrete shells supported only by tangential stresses. In addition, if the mass function satisfies the conditions $m\geqslant 0$, $m'\geqslant 0$ and $r>2m$, then the Weak Energy Condition (WEC)~\cite{Alho:2023ris, Hawking:1973uf} is {\it always} satisfied, since these imply $\rho\geqslant 0$, as well as $p + \rho\geqslant 0$ and $p_T + \rho\geqslant 0$.\\

\begin{figure}[ht]
\includegraphics[width=0.43\textwidth]{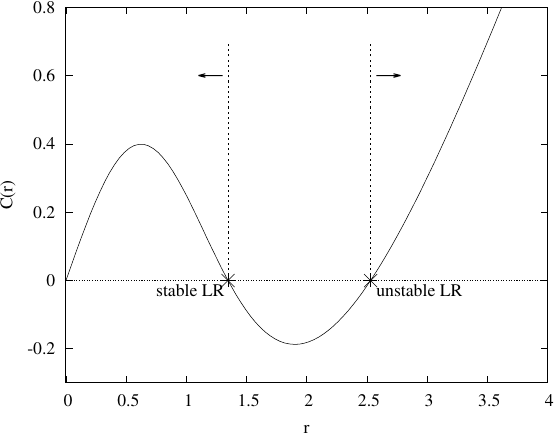}
\caption{\label{fig:LRs} Illustrative plot of the function $C(r)$ defined in~\eqref{eq:LRequation} for the mass function~\eqref{eq:mo_rho}. The roots of $C(r)$ determine the location of the LRs, with the innermost LR being stable and the outermost LR unstable. The vertical dotted lines and arrows represent how small perturbations induced by the stable LR shift the innermost LR inward and the outermost LR outward.}
\end{figure}

{The motion of null geodesics, which is crucial for the results of this paper, is governed by an effective potential $H$, see~\cite{Cunha:2018acu,Cunha:2017qtt}. With no loss in generality, we can restrict the analysis to the equatorial plane ($\theta=\pi/2$),  due to spherical symmetry:}
%================
\begin{equation}
    H(r)\equiv \sqrt{-\frac{g_{tt}}{g_{\varphi\varphi}}}=\sqrt{\frac{e^{-\Psi(r)}}{r^2}\left(1-\frac{2m(r)}{r}\right)}\,.
    \label{eq:Hpotential}
\end{equation}
LRs, or circular photon orbits, are found by solving the condition $H'=0$, which for the metric~\eqref{eq:metric} leads to the compact equation $r=3m(r)$. The solutions of this equation describe the locations of stable or unstable circular orbits around the star. It will be useful to introduce a function $C(r)$, which is zero at the location of the LRs:
%================
\begin{equation}
    C(r) \equiv r-3m(r)\,.
    \label{eq:LRequation}
\end{equation}
It is worth noting that~\eqref{eq:LRequation} is only valid for the metric~\eqref{eq:metric}. Close to the star center, as $r\to 0$, regularity requires $m \sim r^3$, while for large radii the mass function converges to the total star mass $M$. Hence $C(r) \sim r$ in both limits. Due to these boundary conditions, if the star is compact enough to have light rings, any (non-degenerate) zeros of the function $C(r)$ must appear in pairs, as required by well-known LR results for horizonless stars~\cite{Cunha:2017qtt, Cunha:2020azh}. This point is illustrated in Fig.~\ref{fig:LRs}. For the sake of simplicity and clarity, we focus on stars with two distinct (non-degenerate) LRs, although the conclusions do not depend on this assumption.

%###########################
%###########################

Since LRs are defined as the circular orbits of null test particles moving within a fixed background geometry, a natural question arises: what happens when there is a large accumulation of null geodesics very close to a LR orbit? As the {piled-up energy} grows, it may start to backreact on the spacetime itself. Can we make some general statements about how this feedback modifies the properties of LRs? This will be the focus of the next section.  

%%%%%%%%%%%%%%%%%%%%%%%%
\section{Small perturbations induced by energy trapped around a stable light ring}
\label{sec:small_perturbationsLR}
%%%%%%%%%%%%%%%%%%%%%%%%

In order to explore how {the additional energy around the} stable LR affects the star, we will restrict our analysis to {\it static} and {\it spherically symmetric} perturbations, and then examine how they lead to a new equilibrium spacetime configuration for the star. A key goal is to isolate and decouple the pure backreaction of a LR from other effects that may arise due to structural changes in the star's interior. The concept of a continuum-shell star proves useful here. Since such stars are constructed as the continuous limit of many concentric matter shells, each individual star layer can be assumed to be composed of an extremely stiff material, ensuring that it is self-sufficient against gravitational collapse. Under static perturbations, we can impose that the radial position of each layer remains {\it fixed}. This constraint is enforced at the cost of potentially increasing the internal tangential pressures, preventing any significant restructuring of the star while still capturing the backreaction effects of the stable LR. The new mass function under such a perturbation becomes:
%================
\begin{equation}
    m(r) = m_o(r) + \delta m(r)\,,
\end{equation}
where $m_o(r)$ is the original {\it unperturbed} mass function of the star, while $\delta m(r)$ accounts for the perturbation function caused by the additional energy located close to the LR. 

{We can attempt to establish a more precise connection between the accumulation of energy from trapped photons around the stable LR and the perturbation $\delta m$. Since a superposition of multiple photons, in different orbital planes and at the same radius, can be effectively modeled as a single thin spherical {\it photon shell} with a small mass, then a dense collection of photons orbiting a stable LR can be represented as a series of concentric photon shells. How does the presence of these photon shells influence the mass function $m$?  One might initially consider Birkhoff's theorem, but it generally does not apply inside a star, as the region is not vacuum. However, the continuum-shell star model proves useful once again. Since the star itself can be decomposed into multiple thin matter layers separated by vacuum, Birkhoff’s theorem can still be applied in the vacuum regions between the star's matter shells and the photon shells. These photon shells are not necessarily static, and could potentially move over the thin matter layers of the star. The radii of the photon shells could oscillate periodically, expanding and contracting around the stable LR. Despite this motion, Birkhoff's theorem {\it does apply} in the vacuum regions between the shells, even if some are moving.} If we consider a sufficiently large number of such photon shells, each with different oscillation amplitudes and phases, their collective effect can produce a {\it static} contribution $\delta m(r)$ to the mass function (see also~\cite{Andreasson:2015agw, Andreasson:2021lsh}).

Notice that while the mass function $m$ obeys a form of linear superposition, the metric function $\psi$ defined by~\eqref{eq:psiEq} generally does not, which reflects the inherent non-linear nature of General Relativity. 

In the following subsections, we will explore how small perturbations $\delta m$ affect both the location of the light rings (LRs) and the potential $H$.

%%%%%%%%%%%%%%%%%%%%%%%%
\subsection{Shift of the LR location}
\label{sec:shitLR}
%%%%%%%%%%%%%%%%%%%%%%%%

The additional energy near the LR naturally affects some properties of the star, particularly shifting the LR's location. The new LR radius $r_{LR}$ will be given by:  
%================
\begin{equation}
    r_{LR} = r_{LR}^o + {\delta r}_{LR}\,
\end{equation}
where ${\delta r}_{LR}$ represents the radial shift from the original LR radius $r_{LR}^o$. The key question is whether ${\delta r}_{LR}$ and $\delta m$ can be directly related. Since the LR position is determined by the condition $C(r_{LR})=0$, one can write:  
%================
\begin{equation}
    r_{LR}^o + \delta r_{LR} - 3\,m_o\big(r_{LR}^o + \delta r_{LR}\big) - 3\,\delta m_{LR} =0\,,
\end{equation}
where $\delta m_{LR} \equiv \delta m(r_{LR})$. By expanding the $m_o$ function to first order, it leads to:
%================
\begin{equation}
    \delta r_{LR} \simeq \frac{\delta m_{LR}}{1/3 -\left.m_o'\right|_{LR}}\,,
    \label{eq:shiftLR}
\end{equation}
where $\left.m_o'\right|_{LR} \equiv m_o'(r_{LR}^o)$. Provided the reasonable assumption that $\delta m_{LR}>0$, the sign of the $\delta r_{LR}$ is determined by the sign of $\big(1/3-m_o'\big) = C'/3$ at the LR location. A direct inspection of the derivative $C'$ via Fig.~\ref{fig:LRs} then leads to the conclusion that:
%================
\begin{equation}
    \delta r_{LR}^{\textrm{stable}} <0\,\qquad \textrm{and} \qquad  \delta r_{LR}^{\textrm{unstable}}>0
\end{equation}

Thus, the accumulation of additional energy near the LR causes the stable LR to {\it shift inwards}, while the unstable LR is {\it pushed outwards}, as represented in Fig.~\ref{fig:LRs}. This result follows from minimal assumptions, and applies to a wide class of continuum-shell stars.

%%%%%%%%%%%%%%%%%%%%%%%%
\subsection{Modification of the LR potential well}
\label{sec:ModH}
%%%%%%%%%%%%%%%%%%%%%%%%

The perturbation $\delta m$ will also affect the depth of the potential $H$ at the stable LR's location:
\begin{equation}
    H(r_{LR})= H_o(r_{LR}^o) + \delta H_{LR}\,,
\end{equation}
where $H_o(r)$ represents the {\it unperturbed potential function $H$}. Can we make any generic prediction for how this potential is modified? Rather than following the same approach as in the previous subsection, we shall adopt a different method. Since the shift in LR position, $\delta r_{LR}$, scales with $\delta m$ - see Eq.~\eqref{eq:shiftLR} - we introduce a {\it small dimensionless parameter} $h$, where $h\ll 1$. This allows us to express the perturbations in a rescaled form:
%================
\begin{equation}
    r_{LR} = r_{LR}^o + R\,h\,,\qquad m(r) = m_o(r) + \epsilon(r)\,h\,,
\end{equation}
where $R$  is an auxiliary radius, and $\epsilon(r)$ is an auxiliary function, such that $R\,h  = \delta r_{LR}$ and $\epsilon(r)\,h = \delta m(r)$. By enforcing that the functional derivative of the LR condition vanishes with respect to  $h$, we would recover the exact same relation in Eq.~\eqref{eq:shiftLR}:
%================
\begin{equation}
    \left.\frac{d}{dh}C(r_{LR})\right|_{h=0}= 0\,.
\end{equation}
This technique can now be applied to determine how the perturbation $\delta m$ modifies the potential at the stable LR. The variation in $H$ is given by:  
%================
\begin{align}
    \delta H_{LR} &\simeq  h\,\,\left.\frac{d}{dh}H(r_{LR})\right|_{h=0} \\
   % &\\
    &=-H_o(r^o_{LR})\,\,\,\left(\int_{r_{{LR}}^o}^\infty\frac{\delta m(\bar{r})}{\left[\bar{r}-2m_o(\bar{r})\,\right]^2}\,d\bar{r}\right)\,.
    \label{eq:dH}
\end{align}
Finally, under the assumption that $\delta m > 0$ the integral in~\eqref{eq:dH} must be positive definite, and so we conclude that $\delta H_{LR} < 0$. In other words, the perturbation deepens the potential well at the stable LR, making it more pronounced.

%%%%%%%%%%%%%%%%%%%%%%%%
\section{Non-linear perturbative regime}
\label{sec:Non-linear}
%%%%%%%%%%%%%%%%%%%%%%%%

In order to better understand how the presence of a stable LR impacts the properties of a star, we now examine a concrete example. A convenient choice for the unperturbed mass function and corresponding density is:

%================
\begin{equation}
    m_o(r)=\frac{r^3}{3+r^3}\,,\qquad \rho_o(r)=\frac{9}{4\pi (r^3+3)^2}\,.
    \label{eq:mo_rho}
\end{equation}

This mass profile is somewhat arbitrary, but it offers the advantage of allowing some computations to be carried out analytically. Additionally, it satisfies the weak energy condition, since $m_o\geqslant 0$, $m_o'\geqslant 0$ and $r>2m_o$ (see Sec.~\ref{sec:continuum-shell-stars}). The total mass $M$ has been normalized to unity. Notably, this continuum-shell star hosts two LRs. In the following subsections, we shall introduce and analyse two different perturbation models, referred to as Model 1 and Model 2.

%%%%%%%%%%%%%%%%%%%%%%%%
\subsection{Model 1}
\label{sec:model1}
%%%%%%%%%%%%%%%%%%%%%%%%
In this first model, we introduce a {\it localized perturbation} by assuming a Dirac delta distribution for the energy density, centered at the stable LR:
%================
\begin{align}
    &\rho(r) = \rho_o(r) + \left(\frac{\mu}{4\pi\,r^2_{LR}}\right) \,\,\delta(r-r_{LR})\,\\
    %& \\
    &m(r) = m_o(r) + \mu\,\,\Theta(r-r_{LR})\,,
\end{align}

Here, $\Theta(x)$ is the Heaviside step function, and the unperturbed mass and density functions, $m_o(r)$ and $\rho_o(r)$, are given in Eq.~\eqref{eq:mo_rho}. The total mass of the star under this perturbation becomes $m_{\infty}=M=1+\mu$, where  $\mu>0$ is a free parameter controlling the strength of the perturbation. Importantly, $\mu$ is not necessarily small. The quantity $r_{LR}$ represents the radius of the stable LR, determined by the condition $C(r_{LR}) = 0$. Using the property $\Theta(0)=1/2$, we obtain the following 4th-order polynomial equation for $r_{LR}$:
   %================
   \begin{equation}
       2r^4 -r^3 \,(6+3\mu) +6r -9\mu=0\,.
   \end{equation}
This equation has an {\it exact analytical solution}, with the relevant root provided in Appendix~\ref{app:solLR}.  \\

The metric function $\psi(r)$, defined by Eq.~\eqref{eq:psiEq}, simplifies in this model to:
%================
\begin{equation}
    \psi(r) = \int_r^\infty F(\bar{r})\,d\bar{r}\,\,+\,\,\frac{6\mu}{r_{LR}}\,\Theta(r_{LR}-r)\,.
    \label{eq:psiModel1}
\end{equation}
where we have defined:
%================
\begin{equation}
    F(r)\equiv \frac{2\,m_o'(r)}{r-2  m_o(r) -2\mu\,\,\Theta(r-r_{LR})  }\,.
\end{equation}
To further simplify the expressions, we introduce the indefinite integral $I(r)$:
%================
\begin{equation}
    I(r) \equiv \int^r F(\bar{r})\,d\bar{r}\,.
\end{equation}
This integral can also be solved analytically, yielding:
%================
\begin{equation}
    I_{_\lambda}(r) = \log\left(3+r^3\right) -\sum_{k=1}^4 P(a_k)\,\log(r-a_k)\prod_{\substack{ i=1\\ i\neq k}}^4\,(a_k-a_i)^{-1}\,.
\end{equation}
Here $P(x)\equiv 3x^2(x-2-2\mu\,\lambda)$, and the coefficients $a_k\in\mathbb{C}$ above are detailed in Appendix~\ref{app:solInt}. These coefficients depend on both the perturbation parameter $\mu$ and on an index $\lambda\in\{0,1\}$, which acts as a control parameter for $I_{_\lambda}$ and affects both $a_k$ and $P(x)$.
The final expression for the metric function $\psi(r)$ takes the piecewise form:\\
$ $\\
If $r>r_{LR}$:
\begin{equation}
    \hspace{-3.9cm}\psi(r)=-I_{_{\lambda=1}}(r)\,,\\
\end{equation}
If $r=r_{LR}$:
\begin{equation}
    \hspace{-2.4cm}\psi(r)=-I_{_{\lambda=1}}(r_{LR}) +\frac{3\mu}{r_{LR}}\,,\\
\end{equation}
If $r<r_{LR}$:
\begin{equation}
    \hspace{0.8cm}\psi(r)=I_{_{\lambda=0}}(r_{LR}) -I_{_{\lambda=0}}(r) -I_{_{\lambda=1}}(r_{LR}) + \frac{6\mu}{r_{LR}}\,.\\
\label{eq:psiModel1v2}
\end{equation}

%###########################
\begin{figure}[ht]
\includegraphics[width=0.45\textwidth]{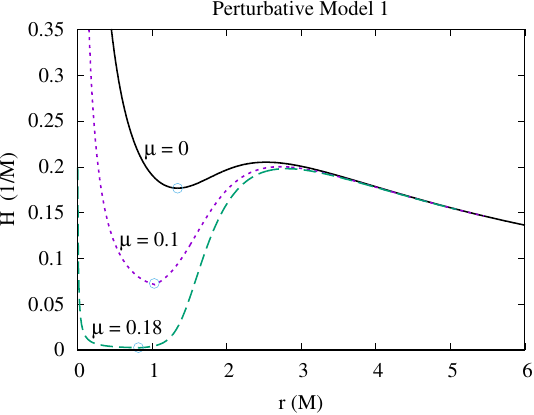}
\caption{\label{fig:Hmodel1} Plot of the geodesic potential $H(r)$ for different values of the perturbation mass scale $\mu$ in Model 1. The unperturbed potential has $\mu=0$. The stable LRs are represented by a blue circle. Both $r$ and $H$ are normalized to the total mass $M$ of the spacetime.}
\end{figure}
%###########################

The geodesic potential $H(r)$, defined in Eq.~\eqref{eq:Hpotential}, is represented in Fig.~\ref{fig:Hmodel1} for different values of $\mu$ in Model 1, with quantities normalized to the total mass $M=1+\mu$. In all cases, the potential has two local extrema:  a maximum corresponding to the unstable LR, and a minimum at the location of the stable LR. From Fig.~\ref{fig:Hmodel1} it is clear that larger perturbations deepen the potential well around the stable LR. Additionally, as $\mu$ increases, the stable LR consistently shifts inward. These trends were anticipated  in Sec.~\ref{sec:shitLR} and Sec.~\ref{sec:ModH}, though only in the regime of small perturbations. Here, however, we extend the analysis to larger perturbations, where non-linear effects can become significant. In Model 1 the value of $\mu$ can be increased up to approximately $\mu\sim 0.2$.  Beyond this threshold the spacetime ceases to be horizonless, making larger values incompatible with the assumed setup. 

Perhaps not surprisingly, both the potential $H(r)$ and the metric function $\psi$ exhibit a discontinuity at $r_{LR}$. However, this discontinuity is not visually apparent from Fig.~\ref{fig:Hmodel1}. This discontinuity arises from the localized {\it delta-like nature} of the perturbation and it is merely an artifact of assuming an idealized delta function. In the next subsection, we shall introduce Model 2, where the perturbation is more spread out, effectively smoothing out this discontinuity. However, this refinement comes at the cost of requiring numerical methods, whereas Model 1 allows a fully analytical treatment. Critically, the results of Model 1 can still be recovered as a limiting case of Model 2, by considering increasingly sharp perturbation distributions.

%%%%%%%%%%%%%%%%%%%%%%%%
\subsection{Model 2}
\label{sec:model2}
%%%%%%%%%%%%%%%%%%%%%%%%
In the second model, we introduce a more {\it dispersed perturbation} by assuming a Gaussian distribution for the energy density, centered at the stable LR:
%================
\begin{equation}
    \rho(r) = \rho_o(r) + \frac{\mu}{\mathcal{A}}e^{-(r-r_o)^2/(2\sigma^2)}\,
    \label{eq:rhoModel2}.
\end{equation}
Here, $\sigma$ represents the standard deviation, controlling the width of the perturbation. Unlike Model 1, where the perturbation was highly localized, this approach spreads the energy more evenly. The corresponding mass function $m(r)$ can be obtained by integrating expression~\eqref{eq:rhoModel2}, though the full analytic expression is lengthy and not particularly illuminating. For details, the reader is referred to Appendix~\ref{app:massM2}. To ensure that the total mass of the perturbed spacetime yields $1+\mu$, we introduced a normalization factor $\mathcal{A}$, with the explicit details also provided in Appendix~\ref{app:massM2}.\\

The Gaussian is initially centered at some radius $r_o$, which we treat as a free parameter. However, to ensure consistency, $r_o$ must ultimately coincide with the location of the stable LR. This adjustment is necessary because the exact position of the LR under the perturbation is not known beforehand. Instead, the new stable LR radius, $r_{LR}$, must be determined numerically by solving the following coupled equations for $(r_{LR}\,,\,r_o)$:  
%================
\begin{equation}
    C(r_{LR}) = 0 \quad \text{and} \quad r_{LR} = r_o\,,
\end{equation}
This approach ensures that the perturbation remains properly centered on the corrected LR position.

%###########################
\begin{figure}[ht]
\includegraphics[width=0.45\textwidth]{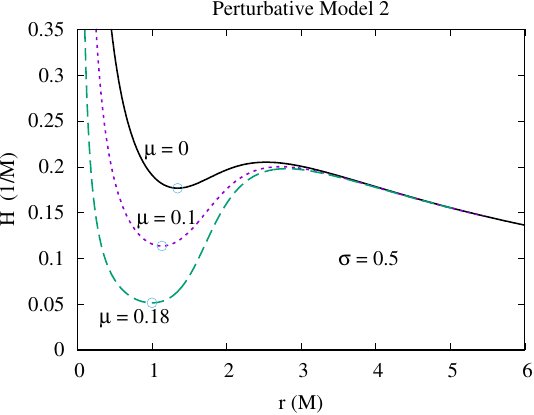}
\caption{\label{fig:Hmodel2} Plot of the geodesic potential $H(r)$ for different values of the perturbation mass scale $\mu$ in Model 2 with $\sigma=0.5$. The unperturbed potential has $\mu=0$. The stable LRs are represented by a blue circle. Both $r$ and $H$ are normalized to the total mass $M$ of the spacetime.}
\end{figure}
%###########################

Figure~\ref{fig:Hmodel2} displays the geodesic potential $H(r)$  for Model 2 with $\sigma=0.5$. To facilitate comparison, we can choose the same values of $\mu$ as in Fig.~\ref{fig:Hmodel1}. While the potentials in Fig.~\ref{fig:Hmodel2} closely resembles those of Fig.~\ref{fig:Hmodel1}, a key difference emerges: in Model 2, the potential-well is not as deep for the same $\mu$. This difference is likely due to the perturbation mass being more spread out in Model 2, compared to the more localized distribution in Model 1. Despite this distinction, both models yield qualitatively very similar results. In fact, as $\sigma\to 0$, Model 2 smoothly transitions into Model 1.

%%%%%%%%%%%%%%%%%%%%%%%%
\section{Beyond continuum-shell stars}
\label{sec:beyond}
%%%%%%%%%%%%%%%%%%%%%%%%

Throughout this article, we have focused on continuum-shell stars: a broad class of stellar models characterized by zero radial pressure ($p = 0$). However, these stars are only a special case within the wider family of static, spherically symmetric spacetimes. How does our analysis change when considering a more general setup with $p\neq 0$?  

Consider the most generic form of a static and spherically symmetric metric:
%================
\begin{equation}
    ds^2 = -e^{-\tilde{\psi}(r)}\left(1-\frac{2m(r)}{r}\right)\,dt^2 + \frac{dr^2}{1-2m(r)/r}+r^2d\Omega^2\,,
\end{equation}
which is structurally similar to Eq.~\eqref{eq:metric}, except for the replacement $\psi\to \tilde{\psi}$, where $\tilde{\psi}$ is now given by:
%================
\begin{equation}
    \tilde{\psi}(r) = \int_r^\infty \frac{2m'(\bar{r})+8\pi\bar{r}^2\,p(\bar{r})}{\bar{r}-2m(\bar{r})}\,d\bar{r}
\end{equation}
Unlike the continuum-shell models, stars described by this metric generally have non-zero radial pressure ($p\neq 0$), meaning that Eq.~\eqref{eq:LRequation} no longer applies. Instead, the locations of the LRs are now determined by the roots of the modified function $\widetilde{C}(r)$:  
%================
\begin{equation}
    \widetilde{C}(r) = r-3m(r)-4\pi r^3\,p(r)\,.
\end{equation}

Similar to the continuum-shell case (Sec.~\ref{sec:continuum-shell-stars}), $\widetilde{C}$ scales as $\sim r$ near the star's center and in the asymptotic limit, assuming the spacetime is asymptotically flat~\footnote{Asymptotic flatness requires $r^2\,p(r)\to 0$ as $r\to \infty$.}. This scaling implies that LRs still appear in pairs (see Fig.~\ref{fig:LRs}), with the stable LR satisfying $\widetilde{C}'<0$. The question now is: how do these locations shift under a small, static, and spherically symmetric perturbation?

Following the approach in Sec.~\ref{sec:ModH}, we introduce perturbations to the mass and pressure profiles, due to the {additional energy around} the stable LR:  
%================
\begin{align}
     &r_{LR} = r_{LR}^o + R\,h\,,\\
     &m(r) = m_o(r) + \epsilon(r)\,h\,,\\
     &p(r) = p_o(r) + P(r)\,h\,,
\end{align}
where $P(r)$ is an auxiliary function characterizing the perturbation in pressure, with $\delta p(r) = h\,P(r)$.  To ensure that the LR condition remains satisfied under these perturbations, we impose that:
%================
\begin{equation}
    \left.\frac{d}{dh}\widetilde{C}\left(r_{LR}\right)\right|_{h=0} =0\,.
\end{equation}
This leads to the following expression for the LR shift:  
%================
\begin{equation}
    \delta r_{LR} =\frac{3\,\delta m_{LR} + 4\pi\,[r^o_{LR}]^3\,\delta p_{LR}}{\widetilde{C}'_{LR}}\,.
    \label{eq:dRGeneric}
\end{equation}
Here, the subscript $LR$ indicates that the quantity is computed at the original LR radius $r^o_{LR}$.\\

To isolate the effects of {the extra energy trapped around the stable LR}, we first assume that the star's matter is extremely stiff. This rigidity ensures that the original matter distribution remains unchanged despite the additional gravitational pull from the extra mass $\delta m$.

If $\delta m>0$, it is natural to expect that the star's interior would respond by increasing its pressure, $\delta p>0$, to maintain equilibrium. This behaviour could follow from some plausible equation of state  $p = p(\rho)$, which relates radial pressure to density. Since the derivative $dp/d\rho$ yields the square of the radial sound speed, it must remain positive to ensure mechanical stability. In other words, it is reasonable to expect that an increase in mass at the LR should naturally be accompanied by an increase in pressure. Let us then assume that both $\delta m_{LR}>0$ and $\delta p_{LR}>0$. Since $\widetilde{C}'_{LR}<0$ at the stable LR location, then Eq.~\eqref{eq:dRGeneric} implies that $\delta r_{LR} < 0$, meaning the stable LR shifts inward. This result is consistent with Sec.~\ref{sec:shitLR}.\\ 

Similarly to Sec.~\ref{sec:ModH}, we can compute how the depth of the potential-well changes under a static perturbation. Considering the modified potential $\tilde{H}(r)$, obtained by the replacement $\psi\to \tilde{\psi}$ in~\eqref{eq:Hpotential}, we obtain:
%================
\begin{equation}
    \delta \tilde{H}_{LR} \simeq -\left(\frac{ \tilde{H}_o(r^o_{LR})}{2}\right)\,\int_{r_{{LR}}^o}^\infty d\bar{r}\,\,\,\mathcal{F}(\bar{r})\,,
\end{equation}
where we have defined 
%================
\begin{equation}
    \mathcal{F}(\bar{r}) \equiv\,\, \frac{8\pi \bar{r}^2\,\delta p(\bar{r})}{\bar{r}-2m_o(\bar{r})} \,+\, \frac{2\,\delta m(\bar{r})\left[1+8\pi\bar{r}^2\,p_o(\bar{r})\right]}{\left[\bar{r}-2m_o(\bar{r})\right]^2}\,.
\end{equation}

If we assume both $\delta m>0$ and $\delta p>0$, then a sufficient condition for the potential to deepen, $\delta\tilde{H}_{LR} <0$, is that the unperturbed (radial) pressure is positive: $p_o>0$. This assumption is well-motivated, as it holds in many physically realistic star models. While it is theoretically possible to construct stars with negative radial pressure, sustained solely by tangential stresses, such configurations are likely to be unstable under non-spherical perturbations. Indeed, a negative radial pressure would likely introduce stronger instabilities than a simple absence of radial pressure  $(p=0)$, which is the case of thin matter shells. The latter have been shown to be generically unstable to non-spherical perturbations in the Newtonian limit~\cite{Bicak:1999tg}.\\

In a more realistic scenario, the behavior of $\delta m$ and $\delta p$ are not necessarily straightforward. If the star's internal matter is allowed to redistribute under the new equilibrium, both $\delta m$ and $\delta p$ could, in principle, take on negative values. This would introduce additional complexity, as the shift in the LR location would then depend on the specific details of the perturbation and how the star's structure adjusts in response. Nevertheless, the fundamental mechanism driving the LR shift should, in principle, persist in more complex scenarios.

%%%%%%%%%%%%%%%%%%%%%%%%
\section{Discussion and Final Remarks}
\label{sec:conclusions}
%%%%%%%%%%%%%%%%%%%%%%%%

In this work, we investigated how perturbations affect the structure of light rings (LRs) in horizonless spacetimes, with a primary focus on continuum-shell stars as a useful toy model. Using both analytical and numerical methods, we demonstrated that perturbations consistently deepen the geodesic potential at the stable LR, while shifting its location inward. This reinforces the idea that the {backreaction of additional energy trapped around} the stable LR can fundamentally modify the spacetime. In contrast to the unstable structures created in~\cite{DiFilippo:2024poc} by {a massive collection of photons around} the unstable LR, the modifications induced by {perturbations around the} {\it stable} LR are not immediately destroyed: if a massive accumulation of photons at the stable LR was perturbed, it would simply oscillate around the stable LR location, as expected from its stability.  

Toward the end of the paper, we have also attempted to extend the analysis to more realistic scenarios, with nonzero radial pressure. {The core conclusions have remained unchanged, provided some additional assumptions.} The consistency of these results across different models strongly suggests that they are an intrinsic feature of the {backreaction of trapped energy around} the stable LR. However, further work is needed to explore how additional internal degrees of freedom of the star might modify these conclusions.\\ 

As a final remark, the plots in Fig.~\ref{fig:Hmodel1} and Fig.~\ref{fig:Hmodel2} bear a striking resemblance to Fig.~6 of Ref.~\cite{Cunha:2022gde}. The latter depicts the null geodesic potential $H$ evolving over time for a Proca star affected by the LR instability. This data, coming from a fully non-linear time evolution of a horizonless Proca star, displays the deepening of the potential-well over time around the stable LR, {at least initially}. This similarity suggests that the nonlinear dynamics in Ref.~\cite{Cunha:2022gde} align with the expectations presented in this work for the {backreaction of perturbations around the} stable LR. Understanding {the latter} is thus essential for uncovering the mechanism driving LR instability, and to whether exotic horizonless spacetimes could be viable black hole mimickers or not.

\newpage
\begin{acknowledgments}

{The author is grateful to Carlos Herdeiro, Eugen Radu and to João Novo for constructive feedback on an earlier version of this manuscript.} The author would also like to thank the Erwin Schrödinger International Institute for Mathematics and Physics (ESI), which provided a stimulating environment for the workshop ``Lensing and Wave Optics in Strong Gravity'', where part of the inspiration for this work took place. P.C. is supported by the Individual CEEC program of 2020 (\cite{funding1}), funded through the Portuguese Foundation for Science and Technology (FCT - Fundação para a Ciência e a Tecnologia)~\cite{FCT_reference}. 
This work is further supported by the Center for Research and Development in Mathematics and Applications (CIDMA) through FCT under the Multi-Annual Financing Program for R\&D Units, 2022.04560.PTDC (\cite{funding2}) and 2024.05617.CERN. 
This work has further been supported by the European Horizon Europe staff exchange (SE) programme HORIZON-MSCA-2021-SE-01 Grant No. NewFunFiCO-101086251.

\end{acknowledgments}

\appendix
\begin{widetext}
%%%%%%%%%%%%%%%%%%%%%%%%
\section{Solution to the LR radius from quartic equation}
\label{app:solLR}
%%%%%%%%%%%%%%%%%%%%%%%%

As detailed in Sec.~\ref{sec:model1}, the stable LR radius $r_{LR}$ is obtained by solving a polynomial equation of the 4th order:
   \begin{equation}
       2r^4 -r^3 \,(6+3\mu) +6r -9\mu=0\,.
   \end{equation}

The relevant root is given by:

\begin{equation}
    r_{LR} = \frac{1}{2}\sqrt{B}  \,\,+\,\,\frac{3}{4}\left(1+\frac{\mu}{2}\right) \,\,-\,\, \frac{1}{2}\sqrt{-B +\frac{27}{4}\left(1+\frac{\mu}{2}\right)^2 +\frac{27\,\left(1+\mu/2\right)^3-24}{4\sqrt{B}}}\,,
\end{equation}
where 
\begin{align}
    &B \equiv \frac{1}{3}\left(\frac{A}{2}\right)^{1/3} +\frac{9}{4}\left(1+\frac{\mu}{2}\right)^2+ 9 \,\left(\frac{2}{A}\right)^{1/3}\left(1-\frac{3\mu}{2}\right)\,,\\
    &A \equiv 243 -\frac{2187}{2}\,\mu\,\left(1+\frac{\mu}{2}\right)^2 + \sqrt{\left[243-\frac{2187}{2}\mu\left(1+\frac{\mu}{2}\right)^2\,\,\right]^2 - 4\,\,\left[27-\frac{81}{2}\mu\right]^3}\,.
\end{align}

    It is worth noting that $A,B\in \mathbb{C}$, while $r_{LR}\in \mathbb{R}$.

%%%%%%%%%%%%%%%%%%%%%%%%
\section{Expression for the coefficients $a_k$}
\label{app:solInt}
%%%%%%%%%%%%%%%%%%%%%%%%

The coefficients $a_k$ with $k\in\{1,2,3,4\}$, introduced in Sec.~\ref{sec:model1} are defined as:
\begin{equation}
    a_k =\alpha_k\frac{\sqrt{\tilde{B}\,\,}}{2}\,+\, \frac{1+\mu\lambda}{2}\,+\,\frac{\beta_k}{2}\sqrt{-\tilde{B} + 3\,\left(1+\mu\lambda\right)^2 + \frac{2\,\alpha_k}{\sqrt{\tilde{B\,}}}\Big[(1+\mu\lambda)^3-3\Big] }\,
\end{equation}
where we have defined:
\begin{align}
    &\tilde{B} \equiv (1+\mu\lambda)^2 \,+\,\frac{1}{3}\left(\frac{\tilde{A}}{2}\right)^{1/3}\,+\, 6\,\left(\frac{2}{\tilde{A}}\right)^{1/3}(1-3\mu\lambda)\,,\\
    &\tilde{A} \equiv 243 \,-\,648\lambda\,\mu\,(1+\mu)^2 \,+\,\sqrt{\Big(243 -648\mu\lambda\,[1+\mu]^2\,\Big)^2 -4\,\Big(18 -54\mu\lambda\Big)^3\,\,}\,,\\
    & \alpha_k= 1-2\,\delta_{k1}-2\delta_{k2}\,,\\
    &\beta_k = 1-2\,\delta_{k1}-2\delta_{k3}\,.
\end{align}
The parameter $\lambda$ can be 1 or 0. One can note that $a_k,\,\tilde{A},\,\tilde{B}\in \mathbb{C}$. However, $I_\lambda(r)\in\mathbb{R}$

%%%%%%%%%%%%%%%%%%%%%%%%
\section{Mass function of Model 2}
\label{app:massM2}
%%%%%%%%%%%%%%%%%%%%%%%%

The mass function $m(r)$ of the perturbed spacetime in Model 2, see~\ref{sec:model2}, is:
\begin{equation}
    m(r) = m_o(r) + \frac{2\sqrt{2\,}\,\sigma\,\pi\mu}{\mathcal{A}}\Biggl\{\sqrt{\pi}\left(\sigma^2 + r_o^2\right)\left[\textrm{Erf}\left(\frac{r-r_o}{\sqrt{2}\,\sigma}\right)+\mathrm{Erf}\left(\frac{r_o}{\sqrt{2}\,\sigma}\right)\right] \,+\,\sqrt{2}\,\sigma\left[r_o\,e^{-r_o^2/(2\sigma^2)} -(r+r_o)\,e^{-(r-r_o)^2/(2\sigma^2)}\right]\Biggl\}\,.
\end{equation}
where $\textrm{Erf}$ is the Gaussian error function. The normalization coefficient $\mathcal{A}$ is given by:
%================
\begin{equation}
    \mathcal{A} = 2\sqrt{2}\sigma\pi\left\{\sqrt{2}\sigma \,r_o \,e^{-r_o^2/(2\sigma^2)} +\sqrt{\pi}\left(\sigma^2 +r_o^2\right)\left[1+ \textrm{Erf}\left(\frac{r_o}{\sqrt{2\,}\,\sigma}\right)\right]\right\}\,,
\end{equation}
This coefficient ensures that in the far-away limit $r\to \infty$, the mass function approaches the total mass $M=1+\mu$.

\end{widetext}

\bibliography{ref}% Produces the bibliography via BibTeX.

\begin{thebibliography}{38}%
\makeatletter
\providecommand \@ifxundefined [1]{%
 \@ifx{#1\undefined}
}%
\providecommand \@ifnum [1]{%
 \ifnum #1\expandafter \@firstoftwo
 \else \expandafter \@secondoftwo
 \fi
}%
\providecommand \@ifx [1]{%
 \ifx #1\expandafter \@firstoftwo
 \else \expandafter \@secondoftwo
 \fi
}%
\providecommand \natexlab [1]{#1}%
\providecommand \enquote  [1]{``#1''}%
\providecommand \bibnamefont  [1]{#1}%
\providecommand \bibfnamefont [1]{#1}%
\providecommand \citenamefont [1]{#1}%
\providecommand \href@noop [0]{\@secondoftwo}%
\providecommand \href [0]{\begingroup \@sanitize@url \@href}%
\providecommand \@href[1]{\@@startlink{#1}\@@href}%
\providecommand \@@href[1]{\endgroup#1\@@endlink}%
\providecommand \@sanitize@url [0]{\catcode `\\12\catcode `\$12\catcode
  `\&12\catcode `\#12\catcode `\^12\catcode `\_12\catcode `\%12\relax}%
\providecommand \@@startlink[1]{}%
\providecommand \@@endlink[0]{}%
\providecommand \url  [0]{\begingroup\@sanitize@url \@url }%
\providecommand \@url [1]{\endgroup\@href {#1}{\urlprefix }}%
\providecommand \urlprefix  [0]{URL }%
\providecommand \Eprint [0]{\href }%
\providecommand \doibase [0]{https://doi.org/}%
\providecommand \selectlanguage [0]{\@gobble}%
\providecommand \bibinfo  [0]{\@secondoftwo}%
\providecommand \bibfield  [0]{\@secondoftwo}%
\providecommand \translation [1]{[#1]}%
\providecommand \BibitemOpen [0]{}%
\providecommand \bibitemStop [0]{}%
\providecommand \bibitemNoStop [0]{.\EOS\space}%
\providecommand \EOS [0]{\spacefactor3000\relax}%
\providecommand \BibitemShut  [1]{\csname bibitem#1\endcsname}%
\let\auto@bib@innerbib\@empty
%</preamble>
\bibitem [{\citenamefont {Kormendy}\ and\ \citenamefont
  {Richstone}(1995)}]{Kormendy:1995er}%
  \BibitemOpen
  \bibfield  {author} {\bibinfo {author} {\bibfnamefont {J.}~\bibnamefont
  {Kormendy}}\ and\ \bibinfo {author} {\bibfnamefont {D.}~\bibnamefont
  {Richstone}},\ }\bibfield  {title} {\bibinfo {title} {{Inward bound: The
  Search for supermassive black holes in galactic nuclei}},\ }\href
  {https://doi.org/10.1146/annurev.aa.33.090195.003053} {\bibfield  {journal}
  {\bibinfo  {journal} {Ann. Rev. Astron. Astrophys.}\ }\textbf {\bibinfo
  {volume} {33}},\ \bibinfo {pages} {581} (\bibinfo {year} {1995})}\BibitemShut
  {NoStop}%
\bibitem [{\citenamefont {Remillard}\ and\ \citenamefont
  {McClintock}(2006)}]{Remillard:2006fc}%
  \BibitemOpen
  \bibfield  {author} {\bibinfo {author} {\bibfnamefont {R.~A.}\ \bibnamefont
  {Remillard}}\ and\ \bibinfo {author} {\bibfnamefont {J.~E.}\ \bibnamefont
  {McClintock}},\ }\bibfield  {title} {\bibinfo {title} {{X-ray Properties of
  Black-Hole Binaries}},\ }\href
  {https://doi.org/10.1146/annurev.astro.44.051905.092532} {\bibfield
  {journal} {\bibinfo  {journal} {Ann. Rev. Astron. Astrophys.}\ }\textbf
  {\bibinfo {volume} {44}},\ \bibinfo {pages} {49} (\bibinfo {year} {2006})},\
  \Eprint {https://arxiv.org/abs/astro-ph/0606352} {arXiv:astro-ph/0606352}
  \BibitemShut {NoStop}%
\bibitem [{\citenamefont {Cardoso}\ and\ \citenamefont
  {Pani}(2019)}]{Cardoso:2019rvt}%
  \BibitemOpen
  \bibfield  {author} {\bibinfo {author} {\bibfnamefont {V.}~\bibnamefont
  {Cardoso}}\ and\ \bibinfo {author} {\bibfnamefont {P.}~\bibnamefont {Pani}},\
  }\bibfield  {title} {\bibinfo {title} {{Testing the nature of dark compact
  objects: a status report}},\ }\href
  {https://doi.org/10.1007/s41114-019-0020-4} {\bibfield  {journal} {\bibinfo
  {journal} {Living Rev. Rel.}\ }\textbf {\bibinfo {volume} {22}},\ \bibinfo
  {pages} {4} (\bibinfo {year} {2019})},\ \Eprint
  {https://arxiv.org/abs/1904.05363} {arXiv:1904.05363 [gr-qc]} \BibitemShut
  {NoStop}%
\bibitem [{\citenamefont {Mazur}\ and\ \citenamefont
  {Mottola}(2023)}]{Mazur:2001fv}%
  \BibitemOpen
  \bibfield  {author} {\bibinfo {author} {\bibfnamefont {P.~O.}\ \bibnamefont
  {Mazur}}\ and\ \bibinfo {author} {\bibfnamefont {E.}~\bibnamefont
  {Mottola}},\ }\bibfield  {title} {\bibinfo {title} {{Gravitational Condensate
  Stars: An Alternative to Black Holes}},\ }\href
  {https://doi.org/10.3390/universe9020088} {\bibfield  {journal} {\bibinfo
  {journal} {Universe}\ }\textbf {\bibinfo {volume} {9}},\ \bibinfo {pages}
  {88} (\bibinfo {year} {2023})},\ \Eprint
  {https://arxiv.org/abs/gr-qc/0109035} {arXiv:gr-qc/0109035} \BibitemShut
  {NoStop}%
\bibitem [{\citenamefont {Herdeiro}\ \emph {et~al.}(2021)\citenamefont
  {Herdeiro}, \citenamefont {Pombo}, \citenamefont {Radu}, \citenamefont
  {Cunha},\ and\ \citenamefont {Sanchis-Gual}}]{Herdeiro:2021lwl}%
  \BibitemOpen
  \bibfield  {author} {\bibinfo {author} {\bibfnamefont {C.~A.~R.}\
  \bibnamefont {Herdeiro}}, \bibinfo {author} {\bibfnamefont {A.~M.}\
  \bibnamefont {Pombo}}, \bibinfo {author} {\bibfnamefont {E.}~\bibnamefont
  {Radu}}, \bibinfo {author} {\bibfnamefont {P.~V.~P.}\ \bibnamefont {Cunha}},\
  and\ \bibinfo {author} {\bibfnamefont {N.}~\bibnamefont {Sanchis-Gual}},\
  }\bibfield  {title} {\bibinfo {title} {{The imitation game: Proca stars that
  can mimic the Schwarzschild shadow}},\ }\href
  {https://doi.org/10.1088/1475-7516/2021/04/051} {\bibfield  {journal}
  {\bibinfo  {journal} {JCAP}\ }\textbf {\bibinfo {volume} {04}},\ \bibinfo
  {pages} {051}},\ \Eprint {https://arxiv.org/abs/2102.01703} {arXiv:2102.01703
  [gr-qc]} \BibitemShut {NoStop}%
\bibitem [{\citenamefont {Sengo}\ \emph {et~al.}(2024)\citenamefont {Sengo},
  \citenamefont {Cunha}, \citenamefont {Herdeiro},\ and\ \citenamefont
  {Radu}}]{Sengo:2024pwk}%
  \BibitemOpen
  \bibfield  {author} {\bibinfo {author} {\bibfnamefont {I.}~\bibnamefont
  {Sengo}}, \bibinfo {author} {\bibfnamefont {P.~V.~P.}\ \bibnamefont {Cunha}},
  \bibinfo {author} {\bibfnamefont {C.~A.~R.}\ \bibnamefont {Herdeiro}},\ and\
  \bibinfo {author} {\bibfnamefont {E.}~\bibnamefont {Radu}},\ }\bibfield
  {title} {\bibinfo {title} {{The imitation game reloaded: effective shadows of
  dynamically robust spinning Proca stars}},\ }\href
  {https://doi.org/10.1088/1475-7516/2024/05/054} {\bibfield  {journal}
  {\bibinfo  {journal} {JCAP}\ }\textbf {\bibinfo {volume} {05}},\ \bibinfo
  {pages} {054}},\ \Eprint {https://arxiv.org/abs/2402.14919} {arXiv:2402.14919
  [gr-qc]} \BibitemShut {NoStop}%
\bibitem [{\citenamefont {Akiyama}\ \emph
  {et~al.}(2019{\natexlab{a}})\citenamefont {Akiyama} \emph
  {et~al.}}]{EventHorizonTelescope:2019dse}%
  \BibitemOpen
  \bibfield  {author} {\bibinfo {author} {\bibfnamefont {K.}~\bibnamefont
  {Akiyama}} \emph {et~al.} (\bibinfo {collaboration} {Event Horizon
  Telescope}),\ }\bibfield  {title} {\bibinfo {title} {{First M87 Event Horizon
  Telescope Results. I. The Shadow of the Supermassive Black Hole}},\ }\href
  {https://doi.org/10.3847/2041-8213/ab0ec7} {\bibfield  {journal} {\bibinfo
  {journal} {Astrophys. J. Lett.}\ }\textbf {\bibinfo {volume} {875}},\
  \bibinfo {pages} {L1} (\bibinfo {year} {2019}{\natexlab{a}})},\ \Eprint
  {https://arxiv.org/abs/1906.11238} {arXiv:1906.11238 [astro-ph.GA]}
  \BibitemShut {NoStop}%
\bibitem [{\citenamefont {Akiyama}\ \emph
  {et~al.}(2019{\natexlab{b}})\citenamefont {Akiyama} \emph
  {et~al.}}]{EventHorizonTelescope:2019ths}%
  \BibitemOpen
  \bibfield  {author} {\bibinfo {author} {\bibfnamefont {K.}~\bibnamefont
  {Akiyama}} \emph {et~al.} (\bibinfo {collaboration} {Event Horizon
  Telescope}),\ }\bibfield  {title} {\bibinfo {title} {{First M87 Event Horizon
  Telescope Results. IV. Imaging the Central Supermassive Black Hole}},\ }\href
  {https://doi.org/10.3847/2041-8213/ab0e85} {\bibfield  {journal} {\bibinfo
  {journal} {Astrophys. J. Lett.}\ }\textbf {\bibinfo {volume} {875}},\
  \bibinfo {pages} {L4} (\bibinfo {year} {2019}{\natexlab{b}})},\ \Eprint
  {https://arxiv.org/abs/1906.11241} {arXiv:1906.11241 [astro-ph.GA]}
  \BibitemShut {NoStop}%
\bibitem [{\citenamefont {Akiyama}\ \emph
  {et~al.}(2022{\natexlab{a}})\citenamefont {Akiyama} \emph
  {et~al.}}]{EventHorizonTelescope:2022wkp}%
  \BibitemOpen
  \bibfield  {author} {\bibinfo {author} {\bibfnamefont {K.}~\bibnamefont
  {Akiyama}} \emph {et~al.} (\bibinfo {collaboration} {Event Horizon
  Telescope}),\ }\bibfield  {title} {\bibinfo {title} {{First Sagittarius A*
  Event Horizon Telescope Results. I. The Shadow of the Supermassive Black Hole
  in the Center of the Milky Way}},\ }\href
  {https://doi.org/10.3847/2041-8213/ac6674} {\bibfield  {journal} {\bibinfo
  {journal} {Astrophys. J. Lett.}\ }\textbf {\bibinfo {volume} {930}},\
  \bibinfo {pages} {L12} (\bibinfo {year} {2022}{\natexlab{a}})},\ \Eprint
  {https://arxiv.org/abs/2311.08680} {arXiv:2311.08680 [astro-ph.HE]}
  \BibitemShut {NoStop}%
\bibitem [{\citenamefont {Akiyama}\ \emph
  {et~al.}(2022{\natexlab{b}})\citenamefont {Akiyama} \emph
  {et~al.}}]{EventHorizonTelescope:2022xqj}%
  \BibitemOpen
  \bibfield  {author} {\bibinfo {author} {\bibfnamefont {K.}~\bibnamefont
  {Akiyama}} \emph {et~al.} (\bibinfo {collaboration} {Event Horizon
  Telescope}),\ }\bibfield  {title} {\bibinfo {title} {{First Sagittarius A*
  Event Horizon Telescope Results. VI. Testing the Black Hole Metric}},\ }\href
  {https://doi.org/10.3847/2041-8213/ac6756} {\bibfield  {journal} {\bibinfo
  {journal} {Astrophys. J. Lett.}\ }\textbf {\bibinfo {volume} {930}},\
  \bibinfo {pages} {L17} (\bibinfo {year} {2022}{\natexlab{b}})},\ \Eprint
  {https://arxiv.org/abs/2311.09484} {arXiv:2311.09484 [astro-ph.HE]}
  \BibitemShut {NoStop}%
\bibitem [{\citenamefont {Goebel}(1972)}]{goebel1972comments}%
  \BibitemOpen
  \bibfield  {author} {\bibinfo {author} {\bibfnamefont {C.}~\bibnamefont
  {Goebel}},\ }\bibfield  {title} {\bibinfo {title} {Comments on the"
  vibrations" of a black hole.},\ }\href@noop {} {\bibfield  {journal}
  {\bibinfo  {journal} {Astrophysical Journal, vol. 172, p. L95}\ }\textbf
  {\bibinfo {volume} {172}},\ \bibinfo {pages} {L95} (\bibinfo {year}
  {1972})}\BibitemShut {NoStop}%
\bibitem [{\citenamefont {Cardoso}\ \emph {et~al.}(2016)\citenamefont
  {Cardoso}, \citenamefont {Franzin},\ and\ \citenamefont
  {Pani}}]{Cardoso:2016rao}%
  \BibitemOpen
  \bibfield  {author} {\bibinfo {author} {\bibfnamefont {V.}~\bibnamefont
  {Cardoso}}, \bibinfo {author} {\bibfnamefont {E.}~\bibnamefont {Franzin}},\
  and\ \bibinfo {author} {\bibfnamefont {P.}~\bibnamefont {Pani}},\ }\bibfield
  {title} {\bibinfo {title} {{Is the gravitational-wave ringdown a probe of the
  event horizon?}},\ }\href {https://doi.org/10.1103/PhysRevLett.116.171101}
  {\bibfield  {journal} {\bibinfo  {journal} {Phys. Rev. Lett.}\ }\textbf
  {\bibinfo {volume} {116}},\ \bibinfo {pages} {171101} (\bibinfo {year}
  {2016})},\ \bibinfo {note} {[Erratum: Phys.Rev.Lett. 117, 089902 (2016)]},\
  \Eprint {https://arxiv.org/abs/1602.07309} {arXiv:1602.07309 [gr-qc]}
  \BibitemShut {NoStop}%
\bibitem [{\citenamefont {McWilliams}(2019)}]{McWilliams:2018ztb}%
  \BibitemOpen
  \bibfield  {author} {\bibinfo {author} {\bibfnamefont {S.~T.}\ \bibnamefont
  {McWilliams}},\ }\bibfield  {title} {\bibinfo {title} {{Analytical Black-Hole
  Binary Merger Waveforms}},\ }\href
  {https://doi.org/10.1103/PhysRevLett.122.191102} {\bibfield  {journal}
  {\bibinfo  {journal} {Phys. Rev. Lett.}\ }\textbf {\bibinfo {volume} {122}},\
  \bibinfo {pages} {191102} (\bibinfo {year} {2019})},\ \Eprint
  {https://arxiv.org/abs/1810.00040} {arXiv:1810.00040 [gr-qc]} \BibitemShut
  {NoStop}%
\bibitem [{\citenamefont {V\"olkel}\ \emph {et~al.}(2022)\citenamefont
  {V\"olkel}, \citenamefont {Franchini}, \citenamefont {Barausse},\ and\
  \citenamefont {Berti}}]{Volkel:2022khh}%
  \BibitemOpen
  \bibfield  {author} {\bibinfo {author} {\bibfnamefont {S.~H.}\ \bibnamefont
  {V\"olkel}}, \bibinfo {author} {\bibfnamefont {N.}~\bibnamefont {Franchini}},
  \bibinfo {author} {\bibfnamefont {E.}~\bibnamefont {Barausse}},\ and\
  \bibinfo {author} {\bibfnamefont {E.}~\bibnamefont {Berti}},\ }\bibfield
  {title} {\bibinfo {title} {{Constraining modifications of black hole
  perturbation potentials near the light ring with quasinormal modes}},\ }\href
  {https://doi.org/10.1103/PhysRevD.106.124036} {\bibfield  {journal} {\bibinfo
   {journal} {Phys. Rev. D}\ }\textbf {\bibinfo {volume} {106}},\ \bibinfo
  {pages} {124036} (\bibinfo {year} {2022})},\ \Eprint
  {https://arxiv.org/abs/2209.10564} {arXiv:2209.10564 [gr-qc]} \BibitemShut
  {NoStop}%
\bibitem [{\citenamefont {Khanna}\ and\ \citenamefont
  {Price}(2017)}]{Khanna:2016yow}%
  \BibitemOpen
  \bibfield  {author} {\bibinfo {author} {\bibfnamefont {G.}~\bibnamefont
  {Khanna}}\ and\ \bibinfo {author} {\bibfnamefont {R.~H.}\ \bibnamefont
  {Price}},\ }\bibfield  {title} {\bibinfo {title} {{Black Hole Ringing,
  Quasinormal Modes, and Light Rings}},\ }\href
  {https://doi.org/10.1103/PhysRevD.95.081501} {\bibfield  {journal} {\bibinfo
  {journal} {Phys. Rev. D}\ }\textbf {\bibinfo {volume} {95}},\ \bibinfo
  {pages} {081501} (\bibinfo {year} {2017})},\ \Eprint
  {https://arxiv.org/abs/1609.00083} {arXiv:1609.00083 [gr-qc]} \BibitemShut
  {NoStop}%
\bibitem [{\citenamefont {Konoplya}\ and\ \citenamefont
  {Stuchl\'\i{}k}(2017)}]{Konoplya:2017wot}%
  \BibitemOpen
  \bibfield  {author} {\bibinfo {author} {\bibfnamefont {R.~A.}\ \bibnamefont
  {Konoplya}}\ and\ \bibinfo {author} {\bibfnamefont {Z.}~\bibnamefont
  {Stuchl\'\i{}k}},\ }\bibfield  {title} {\bibinfo {title} {{Are eikonal
  quasinormal modes linked to the unstable circular null geodesics?}},\ }\href
  {https://doi.org/10.1016/j.physletb.2017.06.015} {\bibfield  {journal}
  {\bibinfo  {journal} {Phys. Lett. B}\ }\textbf {\bibinfo {volume} {771}},\
  \bibinfo {pages} {597} (\bibinfo {year} {2017})},\ \Eprint
  {https://arxiv.org/abs/1705.05928} {arXiv:1705.05928 [gr-qc]} \BibitemShut
  {NoStop}%
\bibitem [{\citenamefont {Glampedakis}\ and\ \citenamefont
  {Pappas}(2018)}]{Glampedakis:2017cgd}%
  \BibitemOpen
  \bibfield  {author} {\bibinfo {author} {\bibfnamefont {K.}~\bibnamefont
  {Glampedakis}}\ and\ \bibinfo {author} {\bibfnamefont {G.}~\bibnamefont
  {Pappas}},\ }\bibfield  {title} {\bibinfo {title} {{How well can ultracompact
  bodies imitate black hole ringdowns?}},\ }\href
  {https://doi.org/10.1103/PhysRevD.97.041502} {\bibfield  {journal} {\bibinfo
  {journal} {Phys. Rev. D}\ }\textbf {\bibinfo {volume} {97}},\ \bibinfo
  {pages} {041502} (\bibinfo {year} {2018})},\ \Eprint
  {https://arxiv.org/abs/1710.02136} {arXiv:1710.02136 [gr-qc]} \BibitemShut
  {NoStop}%
\bibitem [{\citenamefont {Cunha}\ and\ \citenamefont
  {Herdeiro}(2020)}]{Cunha:2020azh}%
  \BibitemOpen
  \bibfield  {author} {\bibinfo {author} {\bibfnamefont {P.~V.~P.}\
  \bibnamefont {Cunha}}\ and\ \bibinfo {author} {\bibfnamefont {C.~A.~R.}\
  \bibnamefont {Herdeiro}},\ }\bibfield  {title} {\bibinfo {title} {{Stationary
  black holes and light rings}},\ }\href
  {https://doi.org/10.1103/PhysRevLett.124.181101} {\bibfield  {journal}
  {\bibinfo  {journal} {Phys. Rev. Lett.}\ }\textbf {\bibinfo {volume} {124}},\
  \bibinfo {pages} {181101} (\bibinfo {year} {2020})},\ \Eprint
  {https://arxiv.org/abs/2003.06445} {arXiv:2003.06445 [gr-qc]} \BibitemShut
  {NoStop}%
\bibitem [{\citenamefont {Cunha}\ \emph {et~al.}(2017)\citenamefont {Cunha},
  \citenamefont {Berti},\ and\ \citenamefont {Herdeiro}}]{Cunha:2017qtt}%
  \BibitemOpen
  \bibfield  {author} {\bibinfo {author} {\bibfnamefont {P.~V.~P.}\
  \bibnamefont {Cunha}}, \bibinfo {author} {\bibfnamefont {E.}~\bibnamefont
  {Berti}},\ and\ \bibinfo {author} {\bibfnamefont {C.~A.~R.}\ \bibnamefont
  {Herdeiro}},\ }\bibfield  {title} {\bibinfo {title} {{Light-Ring Stability
  for Ultracompact Objects}},\ }\href
  {https://doi.org/10.1103/PhysRevLett.119.251102} {\bibfield  {journal}
  {\bibinfo  {journal} {Phys. Rev. Lett.}\ }\textbf {\bibinfo {volume} {119}},\
  \bibinfo {pages} {251102} (\bibinfo {year} {2017})},\ \Eprint
  {https://arxiv.org/abs/1708.04211} {arXiv:1708.04211 [gr-qc]} \BibitemShut
  {NoStop}%
\bibitem [{\citenamefont {Keir}(2016)}]{Keir:2014oka}%
  \BibitemOpen
  \bibfield  {author} {\bibinfo {author} {\bibfnamefont {J.}~\bibnamefont
  {Keir}},\ }\bibfield  {title} {\bibinfo {title} {{Slowly decaying waves on
  spherically symmetric spacetimes and ultracompact neutron stars}},\ }\href
  {https://doi.org/10.1088/0264-9381/33/13/135009} {\bibfield  {journal}
  {\bibinfo  {journal} {Class. Quant. Grav.}\ }\textbf {\bibinfo {volume}
  {33}},\ \bibinfo {pages} {135009} (\bibinfo {year} {2016})},\ \Eprint
  {https://arxiv.org/abs/1404.7036} {arXiv:1404.7036 [gr-qc]} \BibitemShut
  {NoStop}%
\bibitem [{\citenamefont {Cunha}\ \emph {et~al.}(2023)\citenamefont {Cunha},
  \citenamefont {Herdeiro}, \citenamefont {Radu},\ and\ \citenamefont
  {Sanchis-Gual}}]{Cunha:2022gde}%
  \BibitemOpen
  \bibfield  {author} {\bibinfo {author} {\bibfnamefont {P.~V.~P.}\
  \bibnamefont {Cunha}}, \bibinfo {author} {\bibfnamefont {C.}~\bibnamefont
  {Herdeiro}}, \bibinfo {author} {\bibfnamefont {E.}~\bibnamefont {Radu}},\
  and\ \bibinfo {author} {\bibfnamefont {N.}~\bibnamefont {Sanchis-Gual}},\
  }\bibfield  {title} {\bibinfo {title} {{Exotic Compact Objects and the Fate
  of the Light-Ring Instability}},\ }\href
  {https://doi.org/10.1103/PhysRevLett.130.061401} {\bibfield  {journal}
  {\bibinfo  {journal} {Phys. Rev. Lett.}\ }\textbf {\bibinfo {volume} {130}},\
  \bibinfo {pages} {061401} (\bibinfo {year} {2023})},\ \Eprint
  {https://arxiv.org/abs/2207.13713} {arXiv:2207.13713 [gr-qc]} \BibitemShut
  {NoStop}%
\bibitem [{\citenamefont {Benomio}\ \emph {et~al.}(2024)\citenamefont
  {Benomio}, \citenamefont {C\'ardenas-Avenda\~no}, \citenamefont {Pretorius},\
  and\ \citenamefont {Sullivan}}]{Benomio:2024lev}%
  \BibitemOpen
  \bibfield  {author} {\bibinfo {author} {\bibfnamefont {G.}~\bibnamefont
  {Benomio}}, \bibinfo {author} {\bibfnamefont {A.}~\bibnamefont
  {C\'ardenas-Avenda\~no}}, \bibinfo {author} {\bibfnamefont {F.}~\bibnamefont
  {Pretorius}},\ and\ \bibinfo {author} {\bibfnamefont {A.}~\bibnamefont
  {Sullivan}},\ }\bibfield  {title} {\bibinfo {title} {{On turbulence for
  spacetimes with stable trapping}},\ }\href@noop {} {\  (\bibinfo {year}
  {2024})},\ \Eprint {https://arxiv.org/abs/2411.17445} {arXiv:2411.17445
  [gr-qc]} \BibitemShut {NoStop}%
\bibitem [{\citenamefont {Redondo-Yuste}\ and\ \citenamefont
  {C\'ardenas-Avenda\~no}(2025)}]{Redondo-Yuste:2025hlv}%
  \BibitemOpen
  \bibfield  {author} {\bibinfo {author} {\bibfnamefont {J.}~\bibnamefont
  {Redondo-Yuste}}\ and\ \bibinfo {author} {\bibfnamefont {A.}~\bibnamefont
  {C\'ardenas-Avenda\~no}},\ }\bibfield  {title} {\bibinfo {title}
  {{Perturbative and non-linear analyses of gravitational turbulence in
  spacetimes with stable light rings}},\ }\href@noop {} {\bibfield  {journal}
  {\bibinfo  {journal} {Arxiv pre-print}\ } (\bibinfo {year} {2025})},\ \Eprint
  {https://arxiv.org/abs/2502.18643} {arXiv:2502.18643 [gr-qc]} \BibitemShut
  {NoStop}%
\bibitem [{\citenamefont {Di~Filippo}\ and\ \citenamefont
  {Rezzolla}(2025)}]{DiFilippo:2024poc}%
  \BibitemOpen
  \bibfield  {author} {\bibinfo {author} {\bibfnamefont {F.}~\bibnamefont
  {Di~Filippo}}\ and\ \bibinfo {author} {\bibfnamefont {L.}~\bibnamefont
  {Rezzolla}},\ }\bibfield  {title} {\bibinfo {title} {{Can light-rings
  self-gravitate?}},\ }\href {https://doi.org/10.1103/PhysRevD.111.L021504}
  {\bibfield  {journal} {\bibinfo  {journal} {Phys. Rev. D}\ }\textbf {\bibinfo
  {volume} {111}},\ \bibinfo {pages} {L021504} (\bibinfo {year} {2025})},\
  \Eprint {https://arxiv.org/abs/2407.13832} {arXiv:2407.13832 [gr-qc]}
  \BibitemShut {NoStop}%
\bibitem [{\citenamefont {Birkhoff}\ and\ \citenamefont
  {Langer}(1927)}]{birkhoff1927relativity}%
  \BibitemOpen
  \bibfield  {author} {\bibinfo {author} {\bibfnamefont {G.~D.}\ \bibnamefont
  {Birkhoff}}\ and\ \bibinfo {author} {\bibfnamefont {R.~E.}\ \bibnamefont
  {Langer}},\ }\href@noop {} {\emph {\bibinfo {title} {Relativity and modern
  physics}}}\ (\bibinfo  {publisher} {Harvard University Press},\ \bibinfo
  {year} {1927})\BibitemShut {NoStop}%
\bibitem [{\citenamefont {Berry}\ \emph {et~al.}(2022)\citenamefont {Berry},
  \citenamefont {Simpson},\ and\ \citenamefont {Visser}}]{Berry:2022zwv}%
  \BibitemOpen
  \bibfield  {author} {\bibinfo {author} {\bibfnamefont {T.}~\bibnamefont
  {Berry}}, \bibinfo {author} {\bibfnamefont {A.}~\bibnamefont {Simpson}},\
  and\ \bibinfo {author} {\bibfnamefont {M.}~\bibnamefont {Visser}},\
  }\bibfield  {title} {\bibinfo {title} {{General-relativistic thin-shell Dyson
  megaspheres}},\ }\href {https://doi.org/10.1103/PhysRevD.106.084001}
  {\bibfield  {journal} {\bibinfo  {journal} {Phys. Rev. D}\ }\textbf {\bibinfo
  {volume} {106}},\ \bibinfo {pages} {084001} (\bibinfo {year} {2022})},\
  \Eprint {https://arxiv.org/abs/2207.02465} {arXiv:2207.02465 [gr-qc]}
  \BibitemShut {NoStop}%
\bibitem [{\citenamefont {Slav{\'\i}k}(2007)}]{Slavik2007}%
  \BibitemOpen
  \bibfield  {author} {\bibinfo {author} {\bibfnamefont {A.}~\bibnamefont
  {Slav{\'\i}k}},\ }\href@noop {} {\emph {\bibinfo {title} {Product
  integration, its history and applications}}},\ Vol.~\bibinfo {volume} {29}\
  (\bibinfo  {publisher} {Matfyzpress Prague},\ \bibinfo {year}
  {2007})\BibitemShut {NoStop}%
\bibitem [{\citenamefont {Jampolski}\ and\ \citenamefont
  {Rezzolla}(2024)}]{Jampolski:2023xwh}%
  \BibitemOpen
  \bibfield  {author} {\bibinfo {author} {\bibfnamefont {D.}~\bibnamefont
  {Jampolski}}\ and\ \bibinfo {author} {\bibfnamefont {L.}~\bibnamefont
  {Rezzolla}},\ }\bibfield  {title} {\bibinfo {title} {{Nested solutions of
  gravitational condensate stars}},\ }\href
  {https://doi.org/10.1088/1361-6382/ad2317} {\bibfield  {journal} {\bibinfo
  {journal} {Class. Quant. Grav.}\ }\textbf {\bibinfo {volume} {41}},\ \bibinfo
  {pages} {065014} (\bibinfo {year} {2024})},\ \Eprint
  {https://arxiv.org/abs/2310.13946} {arXiv:2310.13946 [gr-qc]} \BibitemShut
  {NoStop}%
\bibitem [{\citenamefont {Alho}\ \emph {et~al.}(2024)\citenamefont {Alho},
  \citenamefont {Nat\'ario}, \citenamefont {Pani},\ and\ \citenamefont
  {Raposo}}]{Alho:2023ris}%
  \BibitemOpen
  \bibfield  {author} {\bibinfo {author} {\bibfnamefont {A.}~\bibnamefont
  {Alho}}, \bibinfo {author} {\bibfnamefont {J.}~\bibnamefont {Nat\'ario}},
  \bibinfo {author} {\bibfnamefont {P.}~\bibnamefont {Pani}},\ and\ \bibinfo
  {author} {\bibfnamefont {G.}~\bibnamefont {Raposo}},\ }\bibfield  {title}
  {\bibinfo {title} {{Spherically symmetric elastic bodies in general
  relativity}},\ }\href {https://doi.org/10.1088/1361-6382/ad1e4b} {\bibfield
  {journal} {\bibinfo  {journal} {Class. Quant. Grav.}\ }\textbf {\bibinfo
  {volume} {41}},\ \bibinfo {pages} {073002} (\bibinfo {year} {2024})},\
  \Eprint {https://arxiv.org/abs/2307.03146} {arXiv:2307.03146 [gr-qc]}
  \BibitemShut {NoStop}%
\bibitem [{\citenamefont {Hawking}\ and\ \citenamefont
  {Ellis}(2023)}]{Hawking:1973uf}%
  \BibitemOpen
  \bibfield  {author} {\bibinfo {author} {\bibfnamefont {S.~W.}\ \bibnamefont
  {Hawking}}\ and\ \bibinfo {author} {\bibfnamefont {G.~F.~R.}\ \bibnamefont
  {Ellis}},\ }\href {https://doi.org/10.1017/9781009253161} {\emph {\bibinfo
  {title} {{The Large Scale Structure of Space-Time}}}},\ Cambridge Monographs
  on Mathematical Physics\ (\bibinfo  {publisher} {Cambridge University
  Press},\ \bibinfo {year} {2023})\BibitemShut {NoStop}%
\bibitem [{\citenamefont {Cunha}\ and\ \citenamefont
  {Herdeiro}(2018)}]{Cunha:2018acu}%
  \BibitemOpen
  \bibfield  {author} {\bibinfo {author} {\bibfnamefont {P.~V.~P.}\
  \bibnamefont {Cunha}}\ and\ \bibinfo {author} {\bibfnamefont {C.~A.~R.}\
  \bibnamefont {Herdeiro}},\ }\bibfield  {title} {\bibinfo {title} {{Shadows
  and strong gravitational lensing: a brief review}},\ }\href
  {https://doi.org/10.1007/s10714-018-2361-9} {\bibfield  {journal} {\bibinfo
  {journal} {Gen. Rel. Grav.}\ }\textbf {\bibinfo {volume} {50}},\ \bibinfo
  {pages} {42} (\bibinfo {year} {2018})},\ \Eprint
  {https://arxiv.org/abs/1801.00860} {arXiv:1801.00860 [gr-qc]} \BibitemShut
  {NoStop}%
\bibitem [{\citenamefont {Andr\'easson}\ \emph {et~al.}(2017)\citenamefont
  {Andr\'easson}, \citenamefont {Fajman},\ and\ \citenamefont
  {Thaller}}]{Andreasson:2015agw}%
  \BibitemOpen
  \bibfield  {author} {\bibinfo {author} {\bibfnamefont {H.}~\bibnamefont
  {Andr\'easson}}, \bibinfo {author} {\bibfnamefont {D.}~\bibnamefont
  {Fajman}},\ and\ \bibinfo {author} {\bibfnamefont {M.}~\bibnamefont
  {Thaller}},\ }\bibfield  {title} {\bibinfo {title} {{Models for
  Self-Gravitating Photon Shells and Geons}},\ }\href
  {https://doi.org/10.1007/s00023-016-0531-4} {\bibfield  {journal} {\bibinfo
  {journal} {Annales Henri Poincare}\ }\textbf {\bibinfo {volume} {18}},\
  \bibinfo {pages} {681} (\bibinfo {year} {2017})},\ \Eprint
  {https://arxiv.org/abs/1511.01290} {arXiv:1511.01290 [gr-qc]} \BibitemShut
  {NoStop}%
\bibitem [{\citenamefont {Andr\'easson}(2021)}]{Andreasson:2021lsh}%
  \BibitemOpen
  \bibfield  {author} {\bibinfo {author} {\bibfnamefont {H.}~\bibnamefont
  {Andr\'easson}},\ }\bibfield  {title} {\bibinfo {title} {{Existence of Steady
  States of the Massless Einstein\textendash{}Vlasov System Surrounding a
  Schwarzschild Black Hole}},\ }\href
  {https://doi.org/10.1007/s00023-021-01104-6} {\bibfield  {journal} {\bibinfo
  {journal} {Annales Henri Poincare}\ }\textbf {\bibinfo {volume} {22}},\
  \bibinfo {pages} {4271} (\bibinfo {year} {2021})},\ \Eprint
  {https://arxiv.org/abs/2102.08170} {arXiv:2102.08170 [gr-qc]} \BibitemShut
  {NoStop}%
\bibitem [{Note1()}]{Note1}%
  \BibitemOpen
  \bibinfo {note} {Asymptotic flatness requires $r^2\protect \,p(r)\to 0$ as
  $r\to \infty $.}\BibitemShut {Stop}%
\bibitem [{\citenamefont {Bicak}\ and\ \citenamefont
  {Schmidt}(1999)}]{Bicak:1999tg}%
  \BibitemOpen
  \bibfield  {author} {\bibinfo {author} {\bibfnamefont {J.}~\bibnamefont
  {Bicak}}\ and\ \bibinfo {author} {\bibfnamefont {B.~G.}\ \bibnamefont
  {Schmidt}},\ }\bibfield  {title} {\bibinfo {title} {{Selfgravitating fluid
  shells and their nonspherical oscillations in Newtonian theory}},\ }\href
  {https://doi.org/10.1086/307560} {\bibfield  {journal} {\bibinfo  {journal}
  {Astrophys. J.}\ }\textbf {\bibinfo {volume} {521}},\ \bibinfo {pages} {708}
  (\bibinfo {year} {1999})},\ \Eprint {https://arxiv.org/abs/gr-qc/9903021}
  {arXiv:gr-qc/9903021} \BibitemShut {NoStop}%
\bibitem [{\citenamefont
  {\href{http://doi.org/10.54499/2020.01411.CEECIND/CP1589/CT0035}{http://doi.org/10.54499/2020.01411.CEECIND/CP1589/CT0035}}()}]{funding1}%
  \BibitemOpen
  \bibfield  {author} {\bibinfo {author} {\bibnamefont
  {\href{http://doi.org/10.54499/2020.01411.CEECIND/CP1589/CT0035}{http://doi.org/10.54499/2020.01411.CEECIND/CP1589/CT0035}}},\
  }\href@noop {} {}\BibitemShut {NoStop}%
\bibitem [{\citenamefont
  {\href{https://ror.org/00snfqn58}{https://ror.org/00snfqn58}}()}]{FCT_reference}%
  \BibitemOpen
  \bibfield  {author} {\bibinfo {author} {\bibnamefont
  {\href{https://ror.org/00snfqn58}{https://ror.org/00snfqn58}}},\ }\href@noop
  {} {}\BibitemShut {NoStop}%
\bibitem [{\citenamefont
  {\href{https://doi.org/10.54499/2022.04560.PTDC}{https://doi.org/10.54499/2022.04560.PTDC}}()}]{funding2}%
  \BibitemOpen
  \bibfield  {author} {\bibinfo {author} {\bibnamefont
  {\href{https://doi.org/10.54499/2022.04560.PTDC}{https://doi.org/10.54499/2022.04560.PTDC}}},\
  }\href@noop {} {}\BibitemShut {NoStop}%
\end{thebibliography}%

\end{document}